\documentstyle[aps,preprint]{revtex}
\begin{document}
\draft
%\preprint{KAIST-CHEP-96-07}
\title{General Static Axially-symmetric Solutions of \\
       (2+1)-dimensional Einstein-Maxwell-Dilaton Theory}
\author{Dahl Park and Jae Kwan Kim}
\address{Department of Physics \\
         KAIST\\
         Taejon 305-701, KOREA }
\date{\today}
\maketitle
\begin{abstract}

We obtain the general static solutions of the axially
symmetric (2+1)-dimensional Einstein-Maxwell-Dilaton theory
by dimensionally reducing it to a 2-dimensional dilaton 
gravity theory.  The solutions consist of the magnetically
charged sector and the electrically charged sector.  We
illuminate the relationship between the two sectors by pointing
out the transformations between them.  

\end{abstract}

\newpage

\section{Introduction}

The low dimensional analogs of the 4-dimensional general relativity 
are useful models to extract the analytic information
about the physics of gravitation, due to their vastly simplified
dynamical content.  Among particularly important models
of this kind are the (2+1)-dimensional general relativity \cite{giddings}
\cite{barrow}
\cite{gott} and the various 2-dimensional dilaton gravity models
\cite{banks}, such as the Callan-Giddings-Harvey-Strominger theory 
\cite{cghs} and the Jackiw-Teitelboim \cite{teitel}
theory.  If we only consider the axially symmetric sector
of the (2+1)-dimensional general relativity, the dynamics of the
problem becomes essentially 2-dimensional.  
The focus of this paper is to obtain the general static
solutions of the axially symmetric (2+1)-dimensional 
Einstein-Maxwell-Dilaton theory by dimensionally reducing it to a 
2-dimensional dilaton gravity theory.

The dimensional reduction we use in this paper has been
originally utilized in \cite{pk} to obtain general static
spherically symmetric solutions of the $D$-dimensional ($D>3$) 
Einstein-Maxwell-Scalar theories.  However, as reported in that
paper, the determination of the analytic solutions of the 3-dimensional 
theory requires a different method, so it was not discussed
there. In this paper, we find it is actually rather
straightforward to give an analytic treatment of the 3-dimensional case.
In that sense this paper is a natural supplement for \cite{pk}.
However, more important motivation for this work comes from the 
fact that the (2+1)-dimensional gravity itself is interesting.
What we get in this paper is general static and axially-symmetric
solutions.  Thus, we recover as special cases the results of \cite{barrow}
where the magnetically charged, static, and axially
symmetric solution is obtained, and \cite{gott}, where one finds
the electrically charged solution.  Moreover we also include a
dilaton field, which plays an important role in string theory,
or other Klein-Kaluza type theories \cite{horowitz}, 
in our general consideration.

In the following section, we present
the dimensional reduction of the (2+1)-dimensional gravity theory
to a 2-dimensional dilaton gravity theory.  The geometrical property 
of the axially-symmetric (2+1)-dimensional space-time is different
from that of the spherically symmetric $D$-dimensional space-time.
Thus, the treatment of $U(1)$ gauge field is quite different from the 
$D$-dimensional Einstein-Maxwell-Scalar theories.  We are led to
separately consider the electrically charged case and the 
magnetically charged case.  The general static solutions of the 
resulting 2-dimensional dilaton gravity theory for electrically
charged case are presented in Appendix, from which
we can also get magnetically
charged solutions by utilizing a electric/magnetic-duality-like
transformation.
Our results in Appendix are in itself interesting, since it gives
exact general static solutions for a class of the 2-dimensional
dilaton gravity theories
not considered elsewhere in literature.  The class of theories
in Appendix contains, as its special case, the (2+1)-dimensional
gravity which is the main concern of this paper.

\section{General Solutions of the (2+1)-dimensional General Relativity}

We consider the axially-symmetric reduction of the (2+1)-dimensional
Einstein-Maxwell-Dilaton theory
\begin{equation}
 I = \int d^3 x \sqrt{g^{(3)}} ( R^{(3)} - \frac{1}{2}
g^{(3) ij} \partial_{i} f \partial_{j} f
+ \frac{1}{4} e^{\chi f} F^2)
\label{daction}
\end{equation} 
to a 2-dimensional dilaton gravity theory.
Here $R^{(3)}$ and $g_{ij}^{(3)}$ represent the 
(2+1)-dimensional scalar curvature and metric tensor, respectively,
and Latin indices $i$, $j$ run over the (2+1) space-time coordinate
labels.
We also have $F$ the curvature 2-form for a $U(1)$ gauge
field and $f$ a dilaton field.  We can write $F_{ij}
= \partial_{i} A_{j} - \partial_{j} A_{i}$
in terms of the vector potential $A_{i}$. The non-zero value of
the real parameter $\chi$ couples the dilaton $f$ to the 
$U(1)$ gauge field in the manner found in the Klein-Kaluza theory
or in the low energy target space effective theory 
of the string theory \cite{horowitz}.
As the first step of the dimensional reduction, we write the
axially symmetric (2+1)-dimensional metric as the sum of
the longitudinal part (with 2-dimensional metric $g_{\alpha \beta}$
where Greek indices $\alpha$, $\beta$ run over the (1+1) space-time
coordinate labels) and the transversal angular part
\begin{equation}
ds^2 = g_{\alpha \beta} dx^{\alpha}dx^{\beta} - 
e^{-4\phi} d \theta^2 . 
\label{metric3}
\end{equation}
The transversal angular part corresponds to a unit circle in
case of the axial symmetry, where $\theta$ corresponds to the 
angle of a point on the circle.  For definiteness, we choose
to describe the longitudinal 2-dimensional space-time in terms
of conformal gauge with conformal coordinate $x^{\pm}$.
In other words, we have $g_{\alpha \beta } dx^{\alpha} dx^{\beta} 
= - \exp (2 \rho ) dx^+ dx^- $ for the longitudinal metric, where
$\exp (2\rho )$ is the conformal factor.
The $\phi$ field, the scale factor
of the transversal metric, will be interpreted as the 2-dimensional
dilaton field under the dimensional reduction.
We use the $(+--)$ signature throughout this paper.
The axial symmetry requires that the metric $g_{\alpha \beta}$, the 
dilaton field $f$ and the 2-dimensional dilaton field $\phi$ 
do not depend on $\theta$.  The three components of the 
(2+1)-dimensional 2-form curvature, $F_{-+}$, $F_{+ \theta}$, and
$F_{- \theta}$ should not depend on $\theta$ either (in fact
their vector potential can also be chosen to be independent of
$\theta$). However, unlike
the higher dimensional spherical symmetry case, $F_{\pm \theta}$ do
not have to vanish in general under the requirement of the axial
symmetry.  

The equations of motion from our action (\ref{daction}) are 
given by
\begin{equation}
R^{(3)}_{ij} - \frac{1}{2} R^{(3)}g^{(3)}_{ij} 
= T^{m}_{ij}
\label{3geo}
\end{equation}
by varying the action with respect to the (2+1)-dimensional
metric tensor,
\begin{equation}
D_{i} \left(e^{\chi f} F^{ij}\right) = 0 
\end{equation}
for the $U(1)$ gauge field, and
\begin{equation}
g^{(3) ij} D_{i} D_{j} f 
 + \frac{\chi}{4} e^{\chi f} F^2 = 0
\end{equation}
for the dilaton field $f$.  Here $D$ denotes the covariant 
derivative in (2+1)-dimensional space-time and 
$T^{m}_{ij}$ is the
stress-energy tensor of the $U(1)$ gauge field and the dilaton
field $f$.  After the imposition of the axial
symmetry, all of the above equations, except the $(\pm , \theta)$
components of Eq. (\ref{3geo}), can be obtained from a
2-dimensional dilaton gravity action
\begin{equation}
I = \int d^2 x \sqrt{-g} e^{-2 \phi} [ R  
-\frac{1}{2} g^{\alpha \beta} \partial_{\alpha} f\partial_{\beta} f 
+ \frac{1}{4} e^{ \chi f } g^{\alpha \beta} g^{\gamma \delta} 
F_{\alpha \gamma} F_{\beta \delta} 
- \frac{1}{2} e^{ \chi f + 4 \phi } g^{\alpha \beta} 
F_{\alpha \theta} F_{\beta \theta} ],
\label{eaction}
\end{equation}
which is obtained from (\ref{daction}) by imposing the axial
symmetry and integrating out the $\theta$ coordinate.  We note
that the sum over the repeated indices run only through $x^+$ and $x^-$,
the 2-dimensional longitudinal space-time.  By varying (\ref{eaction})
with respect to the 2-dimensional metric $g_{\alpha \beta}$, we recover
$(+,+)$, $(-,-)$ and $(-,+)$ component equations of (\ref{3geo}).  The 
$(\theta , \theta )$ component of (\ref{3geo}) is obtained by varying
(\ref{eaction}) with respect to the 2-dimensional dilaton field $\phi$.
The 2-form curvature of the gauge field $F$ is composed of
the electric field $F_{-+}$ and the magnetic field $F_{\pm  \theta}$.
When we consider static and axially symmetric equations of motion,
we can show that $F_{+ \theta } =  \partial_+ A_{\theta } 
= \partial_- A_{\theta} = F_{- \theta }$  since $A_{\pm}$ does
not depend on $\theta$ and, under a suitable choice of conformal
coordinates, $A_{\theta}$ depends only on a space-like coordinate
$x = x^+ + x^-$.  Thus, $F_{\pm \theta}$ contains only magnetic
fields and no electric fields (i.e., $F_{+ \theta} -
F_{- \theta} = 0)$, as far as the static analysis
is concerned.
By inserting (\ref{metric3}) into the $(\pm , \theta )$ components
of the Einstein tensor 
$R^{(3)}_{ij} - g^{(3)}_{ij} R^{(3)} /2$,
we can verify that they identically vanish.  
Consequently, the $(\pm , \theta)$
components of Eq. (\ref{3geo}) reduce to
\[ T^m _{\pm \theta } = \pm e^{\chi f - 2 \rho } F_{+-} 
       F_{\theta \pm} + \frac{1}{2} \partial_{\pm} f 
       \partial_{\theta} f  = 0,\]
which becomes
\begin{equation}
F_{+-} F_{\theta \pm } = 0
\label{cond}
\end{equation}
upon imposing the axial-symmetry.  Thus, our original system 
reduces to a 2-dimensional dilaton gravity action (\ref{eaction}), 
which is of the type we solve in Appendix.  The only missing information
in (\ref{eaction}) is supplied by Eq.(\ref{cond}); it simply 
states that we have either electrically charged solutions or magnetically
charged solutions, but no dyonic solutions which have both magnetic
and electric charges.

First, we consider the purely electrically charged case, for which
we set $F_{\pm \theta} = 0$ and $F_{-+} \ne 0$.  Then, we immediately 
find (\ref{eaction}) reduces to (\ref{oaction}) in Appendix with the
assignment of $\gamma = \mu = \epsilon = 0$.  Thus, we can follow the 
calculations in Appendix leading to the action
\begin{equation}
I_e = \int dx [ \dot{\Omega} \dot{\rho}-\frac{1}{4}\Omega\dot{f}^2
+\frac{1}{4}e^{\chi f-2\rho}\Omega \dot{A}^2] .
\label{ie}
\end{equation}
where we introduce $\Omega = \exp (-2 \phi )$ and a space-like
coordinate $x = x^+ + x^-$.  All the functions depend only on $x$,
and we also introduce $F_{-+} = \dot{A}$ with the overdot
representing the differentiation with respect to $x$.
Getting the general static solution in the conformal gauge
is tantamount to solving the 
equations of motion derived from the action (\ref{ie}) under the gauge
constraint
\begin{equation}
\ddot{\Omega} - 2 \dot{\rho} \dot{\Omega} + 
\frac{1}{2} \Omega \dot{f}^2=0. 
\label{econst}
\end{equation}
>From Eq. (\ref{Aor}), Eq. (\ref{fs}), Eq. (\ref{OaA}) and
Eq. (\ref{Ax}), we get
\begin{eqnarray}
2|Q|e^{2\rho}&=&e^{\chi f_1}e^{2sI(A)} \label{eAor} \\
f(A)&=&\frac{1}{\chi}\left(2sI(A)-\ln{|P(A)|}\right)+f_1 
\label{efs} \\
\Omega^2&=&e^{-4\phi_0} |P(A)|^{-2/\chi^2}
e^{(8s^2-f_2)I(A)/(2s\chi^2)} \label{eOaA}  \\
x-x_0&=& \int \frac{\Omega(A)}{P(A)}dA
\label{eAx}
\end{eqnarray}
where $Q$, $s$, $f_0$, $c$, $f_1$, $\phi_0$ 
and $x_0$ are constants
of integration \cite{note1} and 
\[ P(A) = (2s-\chi f_0) A  +\frac{\chi^2}{2}Q A^2+c .  \]
 Here $Q \ne 0$, 
$f_2=-\chi^2f_0^2+4s\chi f_0+2\chi^2 Qc$ and $I(A)=\int P(A)^{-1}dA$.
>From Eq. (\ref{metrc}), we can rewrite the (2+1)-dimensional metric
in the geometric gauge as
\begin{equation}
ds^2=\frac{P}{2Q}e^{\chi f}
\left[dT^2-\frac{1}{16P^2}\left(\frac{dA}{d\phi} \right)^2 dr^2 \right]
-r^2d\theta^2,
\label{emetrc}
\end{equation}
where $r=\Omega$ and $2T=x^+-x^-$ is the natural time-like coordinate
orthogonal to the space-like coordinate $x= x^+ + x^-$.

Now we consider the purely magnetically charged case where 
$F_{-+} = 0$ and $F_{\pm \theta} \ne 0$.  To obtain the general static
solutions, we once again 
assume all fields depend on a single space-like
variable $x = x^+ + x^-$ under a suitable choice of the conformal
gauge.  The static magnetic field can, thus, be written as $F_{- \theta} 
= F_{+ \theta} = \dot{A}$, where the overdot represents the differentiation
with respect to $x$, as before. Then, the static equations 
of motion from (\ref{eaction})
can be derived from the following action
\begin{equation}
I_m = \int dx [ \dot{\Omega} \dot{\rho} - \frac{1}{4} \Omega \dot{f}^2
     - \frac{1}{4} e^{\chi f}  \Omega^{-1} \dot{A}^2 ] 
\label{im}
\end{equation}
along with the gauge constraint whose static version is given by the
condition
\begin{equation}
\ddot{\Omega} - 2 \dot{\rho} \dot{\Omega} + \frac{1}{2} \Omega
\dot{f}^2 + \frac{1}{2} e^{\chi f}\Omega^{-1} \dot{A}^2 = 0 .
\label{mconst}
\end{equation}
We could follow the analysis in the fashion given
in Appendix to solve the above equations of motion.  However, we take an
alternative route here.  We observe that there exists a transformation 
of fields that maps (\ref{im}) into (\ref{ie}).   Namely, under the 
transformation ${\cal T}_{me}$ defined by
\begin{equation}
 {\cal T}_{me}:(\Omega , e^{\rho} ,  A, f ) \rightarrow 
(e^{\rho} , \Omega,  -iA,  f ),~
 dx          \rightarrow  e^{\rho}\Omega^{-1} dx  , 
\label{dual}
\end{equation}
the magnetic action (\ref{im}) transforms exactly into the
electric action (\ref{ie}).  In addition to this, under
the transformation ${\cal T}_{em}$ defined as
\begin{equation}
{\cal T}_{em}: (\Omega , e^{\rho} ,  A, f ) \rightarrow 
(e^{\rho} , \Omega,  +iA,  f ),~
 dx          \rightarrow  e^{\rho}\Omega^{-1} dx  , 
\label{dual1}
\end{equation}
the electric action (\ref{ie}) changes into the magnetic action 
(\ref{im}).  We note that both ${\cal T}_{em} \circ  {\cal T}_{me} $ and
${\cal T}_{me} \circ {\cal T}_{em}$ are the identity map in the space of
fields.  In fact, recalling that $F_{+-} = \dot{A}$ in the electrically
charged case and $F_{\pm \theta} = \dot{A}$ in the magnetically
charged case, the above transformations are similar to the usual
electric/magnetic duality transformations where one transforms 
$\vec{B} \rightarrow - \vec{E}$ (${\cal T}_{me}$ in our case) and
$\vec{E} \rightarrow \vec{B}$ (${\cal T}_{em}$ in our case).

Given these transformations, it is straightforward to write down
the solutions for the magnetic case utilizing our previous results
for the electrically charged solutions.  We introduce ${\cal T}_{me}$
as a field redefinition
\begin{equation}
\Omega = e^{\bar{\rho}} , \ e^{\rho} = \bar{\Omega}, \  f =\bar{f} ,
\ A = -i \bar{A}, \  d x = e^{\bar{\rho}} \bar{\Omega}^{-1} d \bar{x} . 
\end{equation}
Then the equations of motion for the redefined fields become identical
to those in the electrically charged case and, as a result, the integrated
form of them are given in (\ref{f0})-(\ref{c1}) in Appendix. We have
\begin{eqnarray}
f_0&=&\bar{\Omega}\bar{f}^{\prime}+\frac{\chi}{2}e^{\chi \bar{f}-2\bar{\rho}}
\bar{\Omega}\bar{A}^{\prime} \bar{A} \label{mf0} \\
2iQ&=&e^{\chi \bar{f}-2\bar{\rho}} \bar{\Omega} \bar{A}^{\prime}
\label{ma0} \\
c_0&=&\bar{\rho}^{\prime} \bar{\Omega}^{\prime} - \frac{1}{4}
\bar{\Omega} \bar{f}^{\prime 2} 
+ \frac{1}{4}e^{\chi \bar{f}-2\bar{\rho}}\bar{\Omega} \bar{A}^{\prime 2}
\label{mc0} \\
s+c_0\bar{x}&=& \bar{\rho}^{\prime} \bar{\Omega},
\label{mc1}
\end{eqnarray}
where the prime denotes the differentiation with respect to $\bar{x}$.
The gauge constraint (\ref{mconst}) changes into
\begin{equation}
\bar{\Omega}\bar{\rho}^{\prime \prime}  - \bar{\rho}^{\prime} \bar{\Omega}^{\prime}
+ \frac{1}{2} \bar{\Omega}  \bar{f}^{\prime 2} - \frac{1}{2} e^{\chi \bar{f}
- 2 \bar{\rho} } \bar{\Omega} \bar{A}^{\prime 2}  = 0
\label{tmconst}
\end{equation}
under the field redefinition.  Using $ \bar{\rho}^{\prime \prime} \bar{\Omega}
= c_0- \bar{\rho}^{\prime} \bar{\Omega}^{\prime}$ which is
obtained by differentiating
Eq. (\ref{mc1}), we find that (\ref{tmconst}) precisely reduces to a condition
$c_0 = 0$ just as (\ref{econst}) gives the same condition.    
Thus, we can write down the solutions (using original field variables) 
immediately as follows:
\begin{eqnarray}
2Q \Omega^2 e^{-\chi f}&=&(2s-\chi f_0) A -\frac{\chi^2}{2}Q A^2+c
\equiv P(A) \\
\label{mAor}
f(A)&=&\frac{1}{\chi}\left(2sI(A)-\ln{|P(A)|}\right)+f_1 \\
\label{mfs}
e^{2\rho}&=&e^{-4\phi_0} |P(A)|^{-2/\chi^2}
e^{(8s^2-f_2)I(A)/(2s\chi^2)} \\
\label{mOaA} 
x-x_0&=& \int \frac{\Omega(A)}{P(A)}dA,
\label{mAx}
\end{eqnarray}
where $f_2=-\chi^2f_0^2+4s\chi f_0-2\chi^2 Qc$.  These are the general
static solutions for the magnetically charged case.

The transformations ${\cal T}_{me}$ and ${\cal T}_{em}$ are 
similar to the usual electric/magnetic
duality.  In general time-dependent (2+1)-dimensional case, 
there are two components
for the electric field and a single component for the magnetic field.
Thus, this is not a duality
transformations in the usual sense of the 4-dimensional case.
However, as far as static solutions are concerned, the number
of components for both electric and magnetic fields is one, so the
existence of duality-like transformations is not as bothering
as it first seems.

If $F_{-+} = F_{\pm \theta} = 0$, the equations of motion are solved
in Appendix.  From Eq. (\ref{frho}),
Eq. (\ref{Orho}) and Eq. (\ref{mtrc1}), we have
\begin{eqnarray}
f(z)&=&p \ln|\ln z| + p (\ln |2s| -\rho_0) \\
ds^2&=&(\ln z)^2 dt^2-|\ln z|^{p^2/2}c_1^{-2} z^{-2}(dz^2+z^2 d\theta^2),
\end{eqnarray}
where $p=f_0/s$, $c_1^2=|2s|^{-p^2/2}e^{\rho_0p^2/2-2c_1/s}$,
$\ln{z}=\pm e^{\rho}/(2s)$ and
$dt=\pm 2sdT$. These results are the same as those given in \cite{barrow}.

In the absence of the dilaton field, namely if $f_0=f_1=\chi=0$, our
results reduce to the solutions found in the literature.
For electrically charged solutions, Eq. (\ref{eAor}) becomes
\begin{equation}
2Qe^{2\rho}=2sA+c
\end{equation}
and Eq. (\ref{eOaA}) yields \cite{note2}
\begin{equation}
\Omega=e^{-2\phi_0}e^{-QA/(2s)}.
\end{equation}
The $x$-dependence is given from Eq. (\ref{eAx}) as
\begin{equation}
x-x_0=e^{-2\phi_0}\int \frac{e^{QA/(2s)}}{2sA+c}dA.
\end{equation}
The (2+1)-dimensional metric becomes
\begin{equation}
ds^2= 8Q^2 \ln \left( \frac{r_c}{r} \right) dt^2 -\frac{1}{8Q^2}
\left( \ln \left( \frac{r_c}{r} \right) \right)^{-1}dr^2-r^2 d\theta^2,
\label{got}
\end{equation}
where $\ln r_c =Qc/(4s^2)-2\phi_0$, $r=\Omega$ and $dt=\pm s/(2Q^2)dT$.
The electric field is
\begin{equation}
F_{rt}=\mp \frac{Q}{r}
\end{equation}
The above result is the same as the solution found by Gott {\it et al.}
\cite{gott}.
The magnetically charged solutions are given by carefully taking $\chi
\rightarrow 0$ limit of Eqs. (\ref{mAor})-(\ref{mAx}).
\begin{eqnarray}
2Q\Omega^2&=&2sA+c \\
e^{2\rho}&=&e^{-4\phi_0}e^{QA/s} \\
\frac{dA}{dx}&=&2Q\Omega
\end{eqnarray}
The (2+1)-dimensional metric for magnetic case is
\begin{equation}
ds^2=\frac{e^{-4\phi_0-cQ/(2s^2)}}{4s^2} e^{Q^2r^2/s^2}
\left[dt^2-dr^2 \right]-r^2d\theta^2
\label{ba}
\end{equation}
where $dt=\pm 2sdT$, as given in \cite{barrow}.  We note that even if the metric in
(\ref{ba}) is related to the metric (\ref{got}) via 
${\cal T}_{me} $, the form of each metric looks quite 
different from each other in the geometric gauge.  The choice
of conformal gauge makes the existence of ${\cal T}_{me}$ and
${\cal T}_{em}$ clear.

\section{Discussions}

We get the general axially symmetric static solutions of the
(2+1)-dimensional
Einstein-Maxwell-Dilaton theory in this paper.  
The reason for the difference
between this case and the case of the $D$-dimensional
Einstein-Maxwell-Scalar theories ($D >3$) comes from the difference
of the transversal space in each case.  
In (2+1)-dimensional axially symmetric
geometry, we have an Abelian symmetry, while the rotational symmetries 
for $D>3$ cases are non-Abelian.  Thus, the $s$-wave sector of
the (2+1)-dimensional theory contains the magnetic sector.  
Another indirect consequence of this difference is that the 
decoupled equation for the $\Omega$ field Eq. (\ref{OaA}) is the first 
order differential equation rather than the second order one in 
$D>3$ case.  This makes the analysis in this paper much
simpler than that of \cite{pk}.  This illustrates our general
point that the low dimensional analogs of the 4-dimensional
Einstein theory provide a more analytically tractable framework
for the study of the gravitation.

The recent interest in (2+1)-dimensional gravity is partly
due to \cite{btz} where one finds black hole solutions after
adding the negative cosmological constant term to the gravity
action we consider in this paper \cite{carlip}.  Thus, one of the most
immediate generalizations of this paper is to add the negative
cosmological constant to our action.  It is interesting to note
that the transformations ${\cal T}_{em}$ and ${\cal T}_{me}$
still exchange the magnetic and the electric sector of the
theory even under this generalization, rather similar to the conventional 
electric/magnetic duality in the 4-dimensional Maxwell theory.
Thus, in future attempts to solve the theory following our 
lines, it is sufficient enough to consider only the magnetic
(or electric) sector of the theory.  Additionally,
it remains to be seen whether
one can find dyonic solutions in (2+1)-dimensional
gravity coupled with a $U(1)$ gauge field, once we relax the
condition of the rotational symmetry.  The transformations
${\cal T}_{em}$ and ${\cal T}_{me}$ will be helpful in getting
an answer to this question.

\begin{acknowledgments}
D. Park wishes to thank Y. Kiem for useful discussions.  He also thanks
Prof. D.S. Hwang for carefully reading the manuscript.
\end{acknowledgments}

\appendix

\section{Solutions of a Class of 2-dimensional Dilaton Gravity
Theories}

The action we consider here is given by 
\begin{equation}
I = \int d^2 x \sqrt{-g} e^{-2 \phi} [ R + \gamma g^{\alpha \beta}
\partial_{\alpha} \phi \partial_{\beta} \phi + \mu e^{2\lambda\phi} 
-\frac{1}{2} g^{\alpha \beta} \partial_{\alpha}
f\partial_{\beta} f + \frac{1}{4} e^{\epsilon \phi
+ \chi f } F^2],
\label{oaction}
\end{equation}
where $R$ denotes the 2-dimensional scalar curvature and 
$F$ the curvature 2-form for an
Abelian gauge field. $\phi$ and $f$ represent a dilaton field and 
a massless
scalar field, respectively. The parameters $\gamma$, $\mu$, $\lambda$,
$\epsilon$ and $\chi$ are assumed to be arbitrary real parameters 
satisfying $4-2\lambda-\epsilon=0$ and $4-2\lambda-\gamma+\epsilon=0$. 
This case is not considered in \cite{pk} where they solved
$4-2\lambda-\epsilon \ne 0$ and $4-2\lambda-\gamma+\epsilon=0$ case. 

We choose to work in a conformal gauge given by
$ g_{+-}=-e^{2\rho+\gamma\phi /2}/2,~~~g_{--}=g_{++}=0 $, and choose the
negative signature for a space-like coordinate and the positive signature
for a time-like coordinate.
In this conformal gauge our original action, modulo total derivative terms,
is simplified to be
\begin{equation}
I = \int dx^+ dx^- ( 4 \Omega \partial_+ \partial_- \rho
+ \frac{\mu}{2} e^{2 \rho} \Omega^{-1}
+ \Omega \partial_+ f \partial_- f
 - e^{\chi f-2 \rho } \Omega F_{-+}^2 ),
\label{conaction}
\end{equation}
where $\Omega=e^{-2\phi}$ and $F_{-+}=\partial_- A_+ - \partial_+ A_- $.
     The equations of motion in the conformal gauge are given by
\begin{equation}
\partial_+\partial_-\Omega + \frac{\mu}{4} e^{2 \rho} \Omega^{-1}
 +\frac{1}{2}e^{\chi f-2 \rho } \Omega F_{-+}^2=0,
\label{emrho}
\end{equation}
\begin{equation}
\partial_+\partial_-\rho - \frac{\mu}{8}
\frac{e^{2 \rho}}{\Omega^2} + \frac{1}{4}\Omega
\partial_+ f \partial_- f - \frac{1}{4}
e^{\chi f-2\rho } F_{-+}^2 =0,
\label{emO}
\end{equation}
along with the equations for the massless scalar field
\begin{equation}
\partial_+\Omega\partial_-f + \partial_-\Omega\partial_+f+2\Omega
\partial_+\partial_-f+\chi e^{\chi f-2\rho}\Omega
F_{-+}^2=0,
\label{emf}
\end{equation}
and for the Abelian gauge fields
\begin{equation}
\partial_-(e^{\chi f-2 \rho } \Omega F_{-+})=0,
\end{equation}
\begin{equation}
\partial_+(e^{\chi f-2 \rho } \Omega F_{-+})=0.
\label{emgf}
\end{equation}
The equations for the Abelian gauge fields can be solved to give
\begin{equation}
F_{-+}=e^{-\chi f+2 \rho } \Omega^{-1} Q,
\label{sgauge}
\end{equation}
where $Q$ is a constant.

     To get the solutions we need, in addition to the equations of motion,
the gauge constraints resulting from the choice of the conformal gauge.
They are given by 
\begin{equation}
\frac{\delta I}{\delta g^{\pm \pm}} = 0,
\end{equation}
where $I$ is the original action Eq. (\ref{oaction}).
We obtain the gauge constraints
\begin{equation}
\partial_{\pm}^2 \Omega - 2 \partial_{\pm} \rho \partial_{\pm} \Omega
+ \frac{1}{2} \Omega ( \partial_{\pm} f )^2 = 0.
\label{gconst0}
\end{equation}

     Now we have to find the static solutions of the equations
of motion Eq. (\ref{emrho})-(\ref{emgf}) with the constraints
Eq. (\ref{gconst0}). The general static solutions can be found by
assuming all functions except the gauge field depend on a single
space-like coordinate $x=x^++x^-$. Then from Eq. (\ref{sgauge}) we
observe that the variable $F_{-+}$ automatically becomes dependent
only on $x$, and we can consistently reduce the partial differential
equations into the coupled second order ordinary differential
equations (ODE's). The resulting ODE's except the gauge constraint
can be derived from an effective action
\begin{equation}
I = \int dx [\dot{\Omega} \dot{\rho}- \frac{\mu}{8}
e^{2 \rho}\Omega^{-1}-\frac{1}{4} \Omega\dot{f}^2
+\frac{1}{4} e^{\chi f-2 \rho} \Omega \dot{A}^2],
\label{xaction}
\end{equation}
where the overdot represents taking a derivative with respect to $x$
and $\dot{A}=F_{-+}$.
The gauge constraints become
\begin{equation}
\ddot{\Omega}-2\dot{\rho}\dot{\Omega}+\frac{1}{2}\Omega\dot{f}^2=0.
\label{gconst}
\end{equation}

The general solutions of the above ODE's are the same as the general static
solutions of the original action under a particular choice of the conformal
coordinates.

     The equations of motion can be integrated to the coupled nonlinear first
order ODE by constructing Noether charges of the effective action.
We observe the following four continuous symmetries of the action
Eq. (\ref{xaction}):
\begin{eqnarray*}
&&f \rightarrow f + \alpha, A \rightarrow A e^{-\chi\alpha/2}\\
&&A \rightarrow A + \alpha \\
&&x \rightarrow x +\alpha \\
&&x \rightarrow xe^{\alpha}, \Omega \rightarrow \Omega e^{\alpha},
\end{eqnarray*}
where $\alpha$ is an arbitrary real parameter of each transformation.
The Noether charges for these symmetries are constructed as:
\begin{eqnarray}
f_0&=&\Omega\dot{f}+\frac{\chi}{2}e^{\chi f-2\rho}
\Omega\dot{A}A \label{f0} \\
2Q&=&e^{\chi f-2\rho} \Omega \dot{A}
\label{a0} \\
c_0&=&\dot{\rho}\dot{\Omega} - \frac{1}{4}
\Omega\dot{f}^2 + \frac{\mu}{8}e^{2\rho}\Omega^{-1}
+ \frac{1}{4}e^{\chi f-2\rho}\Omega\dot{A}^2  \label{c0} \\
s+c_0x&=&\dot{\rho}\Omega.
\label{c1}
\end{eqnarray}
Note that the third Noether charge $c_0$ is fixed to be zero 
($c_0 = 0$) by
using the gauge constraint Eq. (\ref{gconst}) and the equation of motion
for $\rho$ which is derived from the effective action Eq. (\ref{xaction}).

First, we find solutions when there is no $U(1)$ gauge field.
In case of $s \neq 0$, using Eq. (\ref{f0}), Eq. (\ref{c1}) and
Eq. (\ref{c0}), we get solutions
for $f$ and $\Omega$ in terms of $\rho$ as
\begin{equation}
f=\frac{f_0}{s}(\rho-\rho_0)
\label{frho}
\end{equation}
\begin{equation}
\Omega=e^{c_1/s}e^{f_0^2(\rho-\rho_0)/(4s^2)}\exp{\left[\mu e^{2\rho}
/(16s^2) \right]} ,
\label{Orho}
\end{equation}
where $\rho_0$ and $c_1$ are constants of integration.
>From Eq. (\ref{c1}) we get $x$-dependence of $\rho$ as
\begin{equation}
x-x_0=s^{-1} \int \Omega(\rho) d\rho ,
\label{rhox}
\end{equation}
where $x_0$ is a constant of integration.
The metric becomes
\begin{equation}
ds^2=4s^2(\ln z)^2dT^2-\frac{\Omega^2}{z^2}(dz^2+z^2 d\theta^2)
\label{mtrc1}
\end{equation}
where $\ln z =\pm e^{\rho}/(2s)$ and $2T=x^+-x^-$.
In case of $s=0$ we get
\begin{eqnarray*}
\rho=\rho_0 \\
f=\int \frac{f_0}{\Omega(x)} dx +f_1 \\
\Omega(x)={\rm arbitrary~function} ,
\end{eqnarray*}
where $\rho_0$ and $f_1$ are constants of integration and $2f_0^2=
\mu e^{2\rho_0}$. The metric becomes
\begin{equation}
ds^2=-e^{2\rho_0} \left[ \frac{1}{4}\left(\frac{d\Omega}{dx}\right)^2
dr^2-dT^2 \right].
\end{equation}
where $r=\Omega$.

Second, we find solutions when the $U(1)$ gauge field does not vanish.
>From Eq. (\ref{f0}), Eq. (\ref{c1}) and Eq. (\ref{a0}) we get
\begin{equation}
2Q e^{2\bar{\rho}}=(2s-\chi f_0) A +\frac{\chi^2}{2}Q A^2+c
\equiv P(A),
\label{Aor}
\end{equation}
where $\bar{\rho}=\rho-\chi f/2$ and $c$ is a constant of
integration.
>From Eq. (\ref{f0}), (\ref{a0}) and (\ref{Aor}) we can determine $f$ via
\begin{equation}
\dot{f}=\frac{f_0-\chi Q A}{P(A)}\dot{A}
\label{fdot}
\end{equation}
which upon integration becomes
\begin{equation}
f(A)=\frac{2s}{\chi}I(A)-\frac{1}{\chi}\ln{|P(A)|}+f_1 ,
\label{fs}
\end{equation}
where $I(A)=\int P(A)^{-1}dA$ and $f_1$ is a constant of integration.
The constant of integration $f_1$ represents
the trivial constant term which we can add to the scalar field $f$.
Using Eq. ({\ref{a0}) and (\ref{Aor}), we can rewrite
Eq. (\ref{c0}) as
\begin{equation}
8s\frac{d\phi}{dA}=\frac{4sQA+2Qc-f_0^2}{P(A)}+\frac{\mu}{4Q}e^{\chi f}.
\label{OaA}
\end{equation}
By integrating the above equation we get $\Omega(A)$ as a function of $A$.
Since $Q$ does not vanish, we can find $A$ as a function of $x$ by plugging
Eq. (\ref{Aor}) into Eq. (\ref{a0}),
\begin{equation}
x-x_0= \int \frac{\Omega(A)}{P(A)}dA,
\label{Ax}
\end{equation}
where $x_0$ is the constant of integration. The metric is given by
\begin{equation}
ds^2=-\frac{P}{2Q}e^{\chi f-\gamma \phi /2}
\left[\frac{1}{16P^2}\left(\frac{dA}{d\phi} \right)^2 dr^2-dT^2 \right].
\label{metrc}
\end{equation}


\begin{thebibliography}{99}
\bibitem{giddings} S. Giddings, J. Abbott and K. Kuchar,
                   Gen. Rel. and Grav. {\bf 16}, 751 (1984).
\bibitem{barrow} J. D. Barrow, A. B. Burd and D. Lancaster,
                 Class. Quantum Grav. {\bf 3}, 551 (1986).
\bibitem{gott} J. R. Gott, III, J. Z. Simon and M. Alpert, Gen. Rel. and
               Grav. {\bf 18}, 1019 (1986).
\bibitem{banks} T. Banks and M. O'loughlin, Nucl. Phys.
                {\bf B362}, 649 (1991); R.B. Mann,
                Gen. Rel. Grav. {\bf 24}, 433 (1992).
\bibitem{cghs} C.G. Callan, S.B. Giddings, Harvey and A. Strominger,
               Phys. Rev. {\bf D45}, R1005 (1992).
\bibitem{teitel} C. Teitelboim, in {\it Quantum Theory of Gravity},
                 S. M. Christensen, ed. (Adam Hilger, Bristol, 1984);
                 T. Banks and O'Loughlin, Nucl. Phys. {\bf B362}, 649 (1991).
\bibitem{pk} D. Park and Y. Kiem, Phys. Rev. {\bf D53}, 5513 (1996).
\bibitem{horowitz} G. Horowitz, in {\ String Theory and Quantum Gravity '92, 
                   Proceedings of the Trieste
                   Spring School and Workshop}, J. Harvey et. al., ed.,
                   (World Scientific, 1993) and references cited therein. 
\bibitem{note1} We may naively expect there should be
                8 constants of integration.  However, the gauge constraint 
                sets one constant of motion to be zero.
\bibitem{note2} One should be very careful in
                taking $\chi \rightarrow 0$ limit of Eq. (\ref{efs}) and
                Eq. (\ref{eOaA}).
\bibitem{btz} M. Banados, C. Teitelboim, J. Zanelli, 
              Phys. Rev. Lett. {\bf 69}, 1849 (1992);
              M. Banados, M. Henneaux, C. Teitelboim, J. Zanelli,
              Phys. Rev. {\bf D48}, 1506 (1993).
\bibitem{carlip} S. Carlip, Class. Quantum Grav. {\bf 12}, 2853 (1995).


\end{thebibliography}
\end{document}